# Spicules Intensity Oscillations in SOT/HINODE Observations

E. Tavabi[1], A. Ajabshirizadeh[2,3], A. R. Ahangarzadeh Maralani[2,3] and S. Zeighami[2,3]

1- Physics Department, Payame Noor University (PNU), 19395-3697-Tehran, I. R. of Iran,

2- Center for Excellence in Astronomy & Astrophysics (CEAA), Research Institute for Astronomy & Astrophysics of Maragha (RIAAM), Maragha, Iran, P.O.Box: 55134-441,

3- Department of Physics, Tabriz Branch, Islamic Azad University, Tabriz, Iran.

(Emails: tavabi@iap.fr)

**Abstract**

Aims. We study the coherency of solar spicules intensity oscillations with increasing height above the solar limb in quiet Sun, active Sun and active region using observations from HINODE/SOT. Existence of coherency up to transition region strengthens the theory of the coronal heating and solar wind through energy transport and photospheric oscillations.

Methods. Using time sequences from the HINODE/SOT in Ca II H line, we investigate oscillations found in intensity profiles at different heights above the solar limb. We use the Fourier and wavelet analysis to measure dominant frequency peaks of intensity at the heights, and phase difference between oscillations at two certain heights, to find evidence for the coherency of the oscillations.

Results. The results of fast Fourier transform (FFT) for the quiet Sun, active Sun and active region show frequency peaks of order 3.6 mHz, 5.5 mHz and 7.3 mHz at four separate heights. The fluctuations of powers are randomly for the three datasets, i.e., independent from height and solar activity. The wavelet results for quiet Sun, active Sun and active region indicate dominant frequencies as similar to FFT results. Results of coherency represent frequencies at about 3.5 mHz and 5.5 mHz for all three datasets. Histograms of frequencies corresponding to maximum coherency for quiet Sun , active region and active sun display frequencies about 3.5 mHz, 4.2 mHz, 4.6 mHz ,5.3 mHz and 5.8 mHz. The phase speeds of 50-450 km/s are measured for quiet Sun, 50-560 $kms^{-1}$ for active region and 50-550 km/s for active Sun. The majority of the measured phase speeds in locations where there is known to be considerable dynamic activity are more than quiet Sun, and the phase speeds obtained from three datasets increase with height. We find also strong evidence for upwardly propagating waves with high coherency in three datasets. Intensity oscillations may result from the presence of the coherent waves, which could provide significant energy to heat the solar atmosphere. Finally, we can calculate the energy and the mass transported by spicules providing energy equilibrium, according to density values of spicules at different heights. To extend this work, we can also consider coherent oscillations at different latitudes and suggest to study of oscillations which may be obtained from observations of other satellites.

## 1. Introduction

The solar chromosphere is one of the most dynamic layers of solar atmosphere which can be observed by chromospheric rather cool lines (Hα and Ca II lines). Spicules are bright jet-like structures that can be seen at the solar limb at the rosettes of network. Some spicules have downward motions due to gravitational force while the others fade off after reaching the maximum height. Bright/dark mottles on disk are equivalent of spicules, but it is not clearly known bright mottles actually correspond to spicules.

A simple cause of spicule formation can be energy release and pressure enhancement in the photosphere or the chromosphere. The pressure enhancement in the deeper layer can be caused a slow mode wave that propagated upward along a flux tube. The slow mode waves suddenly form shocks and the slow shocks can push up the transition region, resulting in the ejection of the chromospheric materials seen as spicules (Suematsu et al. 1982; Shibata & Suematsu 1982). Shibata & Suematsu (1982) studied spicules due to pressure enhancement above the middle chromosphere.

The photospheric convection motions drive Alfvén waves formed shocks to push up the transition region (Hollweg et al. 1982; Kudoh & Shibata 1999). Suematsu (1990); DePontieu et al. (2004) and Hansteen et al. (2006) represented that the wave energy for spicule formation rise from p-mode leakage. Steiner et al. (1998) and Takeuchi (1999) reported spicule formation is because of convective collapse and Matsumoto & Shibata (2010) suggested that the cause is from horizontal granular motions. The magnetic reconnection is also important mechanisms for spicule formation (Suematsu 1998).

DePontieu et al. (2007b) proposed that there are at least two types of spicules, type I spicules or classical spicules go up from the limb and then fall back with 20 $kms^{-1}$ velocities and 3–7 min lifetimes, type II spicules only go upward with ~100 $kms^{-1}$ velocities, 50-100 s lifetimes. Sterling et al. (2010) examined spicules at and near the limb in a polar coronal hole that apparently fall in the class of type II spicules. Based on the spicules diameter Tavabi et al. (2011) have recently reported that there are four types of spicules ranging from 220, 360, 550 to 850 km.

Temperature of the solar atmosphere increases up to 1 MK from transition region (TR). Cause of this rapid rising of the temperature is still unresolved. An idea for this problem could be energy transport through the convective motions and solar oscillations that may be exited magneto-acoustic waves. More of the brightening from the chromosphere originates in spicules. Propagation of the waves can carry applicable energy into the corona. The energy transport operation can be followed by observational evidence of the oscillatory phenomena in the chromosphere (Zaqarashvili et al. 2010).

Tavabi et al. (2014) found surge-like behavior of solar polar region spicules supports the untwisting multicomponent interpretation of spicules exhibiting helical dynamics. Several tall spicules are found with upward and downward flows that are similar at lower and middle levels, the rate of upward motion being slightly higher at high levels; the left and right-hand velocities are also increasing with height; a large number of multicomponent spicules show shearing motion of both left and right-handed senses occurring simultaneously, which might be understood as twisting (or untwisting) threads. The number of turns depends on the overall diameter of the structure made of components and changes from at least one turn for the smallest structure to at most two or three turns for surge-like broad structures. They have already demonstrated that transverse waves observed in solar spicules are adequately described in terms of kink modes of a straight magnetic cylinder embedded into a magnetic environment.

Tavabi et al. (2014) analyzed in detail the proper transverse motions of mature and tall polar region spicules for different heights. Assuming that there might be Helical-Kink waves or Alfvénic waves propagating inside their multi-component substructure, by interpreting the quasi-coherent behavior of all visible components presumably confined by a surrounding magnetic envelop. In addition, they concentrated the analysis on the taller Ca II spicules more relevant for coronal heights and easier to measure.

Tavabi (2014) studied off-limb and on-disk spicules to find a counterpart of the limb spicule on the disk and found a definite signature with strong power in the 3 min (5.5 mHz) and 5 min (3.5 mHz), a full range of oscillations and the high frequency intensity fluctuation (greater than 10 mHz or less than 100 s) corresponding to the occurrence of the so-called type II spicules.

A magnetic flux tube supports several kinds of magnetic body waves, namely torsional waves and transverse and longitudinal magneto-acoustic waves (the latter two are also called kink and sausage waves, respectively). The torsional waves (Alfvén waves) are non-dispersive and propagate for any frequency. Banerjee et al. (2001) exhibited that the network oscillations are bursty and periodic in nature with a finite life time of ~10-20 min. Judge et al. (1997), Hansteen et al. (2000) have reported on coherent oscillations at networks bright pointes.

Gupta et al. (2013) detected from the chromosphere to the transition region period bands 2–4 min, 4–6 min, 6–15 min, and long–period oscillations with periods between 15 and 30 min in bright magnetic points. He et al. (2009) found that spicules are modulated by high-frequency (≥0.02 Hz) transverse fluctuations that propagate upward along the spicules with phase speed ranges from 50 to 150 $kms^{-1}$ and some of the modulated spicules show clear wave-like shapes with short wavelengths less than 8 Mm.

In this work, we use three time sequences observed by HINODE/SOT in Ca II H (396.8nm). We obtain the frequency of oscillations by FFT and wavelet analysis at different heights and then we investigate the coherency of oscillations with phase difference among different heights.

## 2. Observations

We use three time sequences of the quiet Sun, active Sun and active region from equator of solar limb observed in Ca II H line by HINODE/SOT (the wavelength pass-band being centered at 396.8 nm with a FWHM of 0.3 nm). A description of the observations is summarized in Table 1.

**Table 1** The datasets obtained by HINODE/SOT in Ca II H Line.

| Date | 22 November 2006 | 26 January 2007 | 15 January 2014 |
|---|---|---|---|
| Type of Activity | Quiet Sun | Active Region | Active Sun |
| Image size (pixels$^2$) | 1024×1024 | 1024×1024 | 1024×1024 |
| Start (UT) | 00:00:04 | 06:16:10 | 07:01:33 |
| End (UT) | 02:13:03 | 09:53:58 | 11:59:36 |
| Pixel size (arcsecs) | 0.05448 (~40 km) | 0.05448 (~40 km) | 0.10896 (~80 km) |
| Pointing X (arcsecs) | 960.188 | 966.188 | 966.970 |
| Pointing Y (arcsecs) | -89.993 | -71.258 | -171.87 |
| X-FOV (arcsecs) | 55.787 | 55.787 | 111.575 |
| Y-FOV (arcsecs) | 55.787 | 55.787 | 111.757 |
| Cadence (s) | 8 | 8 | 30 |
| Exposure time (s) | 0.51 | 0.21 | 0.31 |
| Number of frames | 1000 | 900 | 600 |
| Sample of the image | figure 1 panel a | figure 1 panel c | figure 1 panel e |

We used the standard SOT subroutines (WV_DENOISE, UNSHARP_MASK, FG_SHIFT_PIX and FG_BAD_PIX) for calibration of raw (level 0) data. The subroutines can be found in the SSWIDL software. These subroutines correct the CCD readout anomalies, bad pixels and flat-field; subtract the dark pedestal. The SOT routine FG_PREP reduces the image spikes, jitter, and aligns the time series.

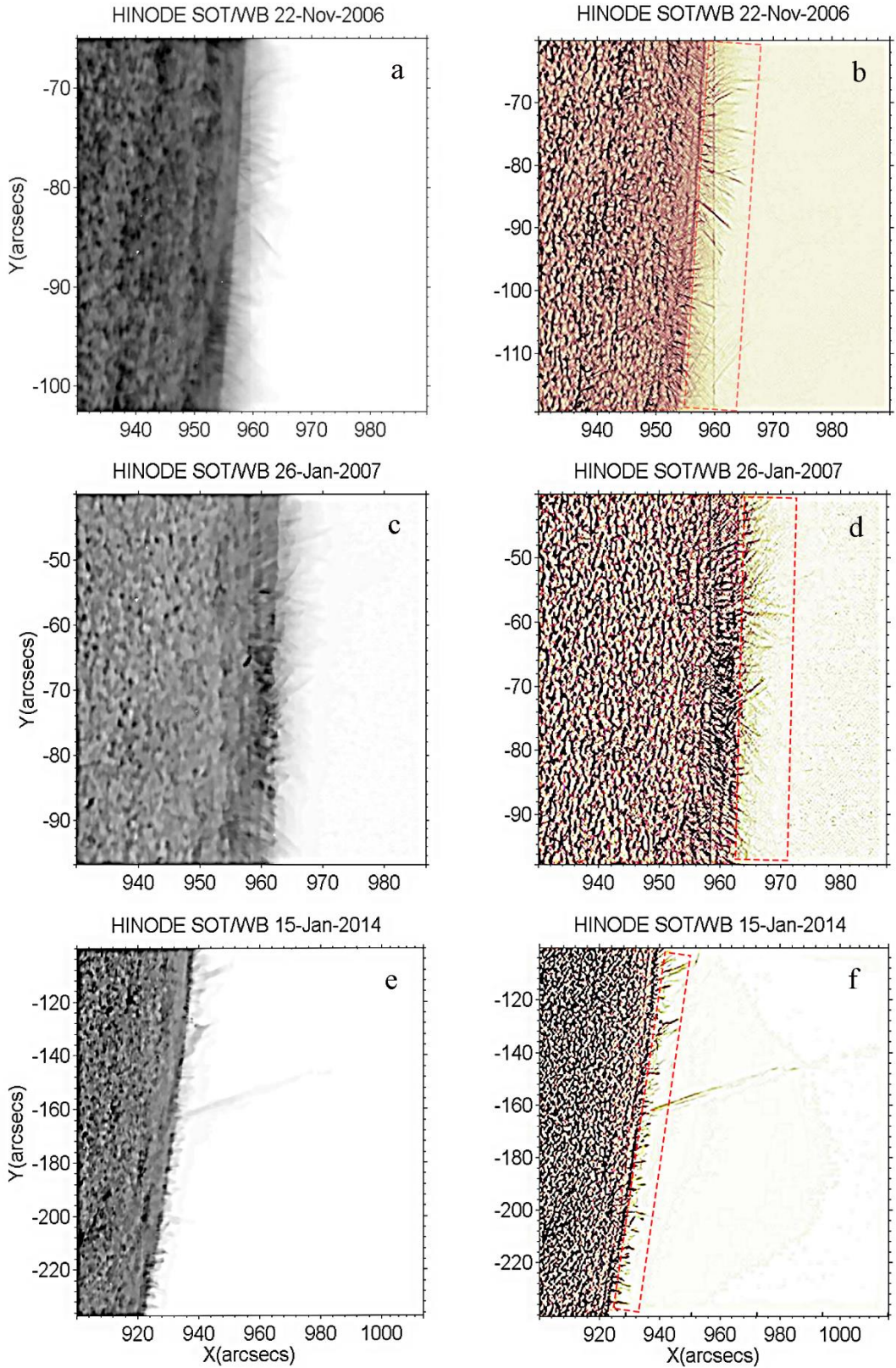

**Fig. 1** Left panels (a, c, e): Negative images obtained by SOT in Ca II H (396.8nm). Right panels (b, d, f): Processed negative images to improve the visibility of fine features after applying the mad-max operator on SOT (HINODE) filtergrams in Ca II H line, obtained on 22 November 2006 (quiet Sun), 26 January 2007 (active region) and 15 January 2014 (active Sun) near the equator.

**3. Data analysis**

We applied mad-max operator to clear the images as shown in figures 1b, 1d and 1f. The operator reduces considerably background noise (Koutchmy & koutchmy 1989; November & koutchmy 1996; Tavabi et al. 2013 and Tavabi 2014). This non-linear spatial filtering clearly shows relatively rather bright radial threads in the chromosphere as fine as the resolution limit of about 120 km.

To study the spatial variations of spicules at different times it is needed that we prepare time-slices of the processed images in the Y direction. The algorithm is as follows located a specific row from all the images (Tavabi 2014), See figure 2. A drift speed due to instrumental effect toward the east was identified from solar limb motion in the panels of figure 2. The drift for 22 November 2006 is an average speed less than 0.015"/min, for 26 January 2007 is 0.021"/min and for 15 January 2014 is 0.017"/min.

We can plot intensity versus time (intensity profiles) as illustrated in the above panel of wavelet spectra of figures 6, 7 and 8 and calculate parameters for example coherency, time delay, phase difference and phase speed by Fourier and Wavelet Transforms.

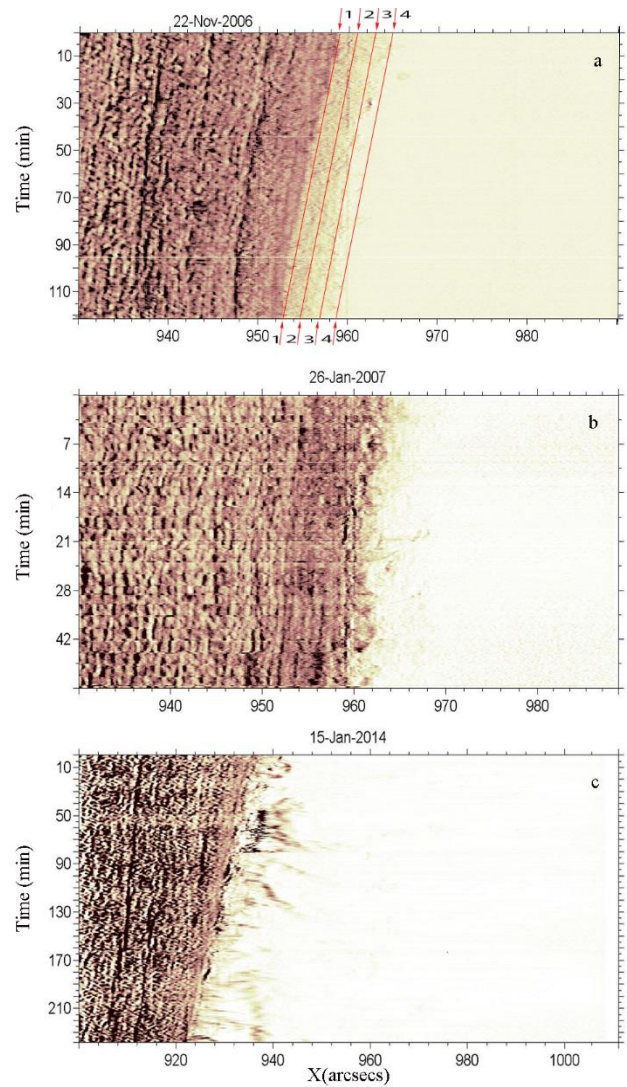

**Fig. 2** Time-Y cut diagrams in negative for the three datasets. Panels a, b, c represent negative time–slice of image obtained on 22 November 2006, 26 January 2007 and 15 January 2014, respectively. The positions of four separate heights which were selected for analysis are shown by four red color arrows (1, 2, 3 and 4) on panel a, at a distance 35 pixels from each other for 22 November 2006 and 26 January 2007 datasets. It is noted that the distance is 17 pixels from each other for 15 January 2014. A drift speed toward the east was identified from solar limb motion in above pictures. The drift for 22 November 2006 is an average speed less than 0.015"/min , for 26 January 2007 is 0.021"/min and for 15 January 2014 is 0.017"/min.

We obtain the intensity profiles for four separate heights and the data is stored. The four separate heights are shown in figure 2 panel a, for 22 November 2006 data by1, 2, 3 and 4 red color arrows. The heights start from solar limb to 105 pixels (at a distance 35 pixels from each other, respectively) out of limb equivalent to 4200 km for 22 November 2006 and 26 January 2007 datasets, and 51 pixels (at a distance 17 pixels from each other, respectively) equivalent to 4200 km for 15 January 2014 dataset. In order to reduce noise effects we average over five neighbor pixels in the intensity profiles at four separate heights.

For further review, we used the Fourier and wavelet transforms on the intensity profiles to obtain the dominant frequencies of oscillations (see figures 6, 7 and 8). Then we run the phase-difference analysis at four separate heights. We calculate the speed of oscillations at different heights from the time delay results.

### 4. Results and discussion

#### 4.1. FFT in heights

We used FFT analysis to obtain power spectra of intensity profiles. The results could be seen in figures 3, 4 and 5 for three time series. We used smoothing and plot the Fourier power above 0.75% level, for the figures, therefore the plots are sharp.

The result of FFT plots for the quiet Sun, active Sun and active region show frequencies about 3.6 mHz, 5.5 mHz and 7.3 mHz at four separate heights. It can be seen that increase and reduction of powers are not regular at four separate heights in the three datasets, more like randomly behavior. It is noted that the results randomly are obtained for selective time-slices, and the results consist which obtained by Hanstoen et al. (2000) and Tavabi (2014) and Gupta et al. (2013).

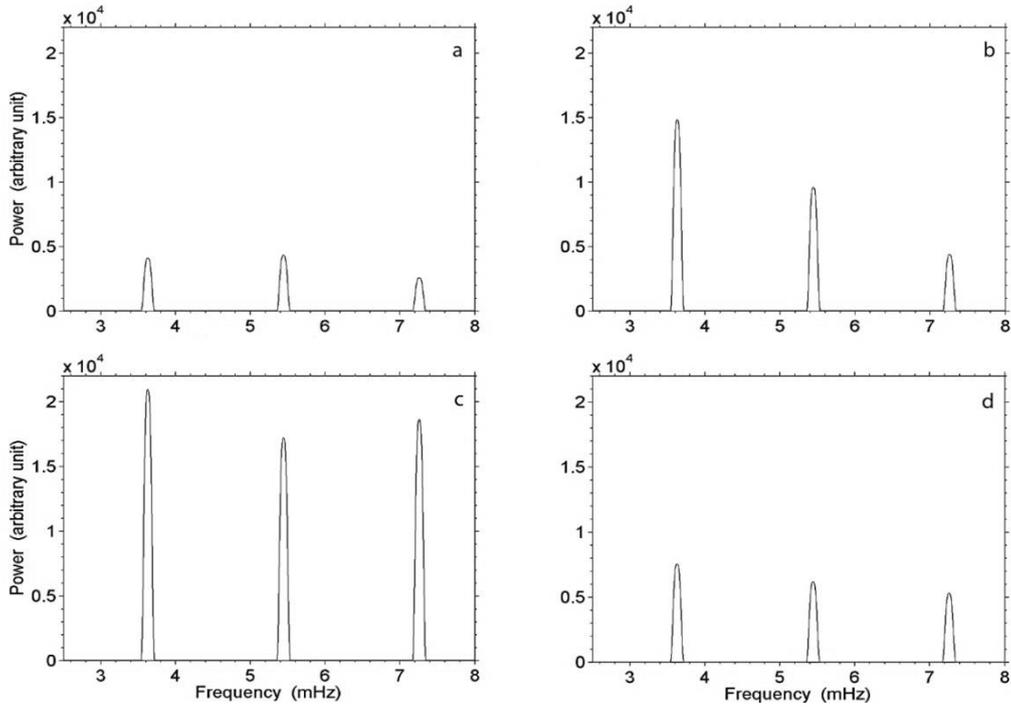

**Fig. 3** Fourier power spectrums obtained on 22 November 2006. The panels a, b, c and d are shown power spectrums, correspond to four separate heights labeled by four red color arrows, respectively, in figure 2 panel a.

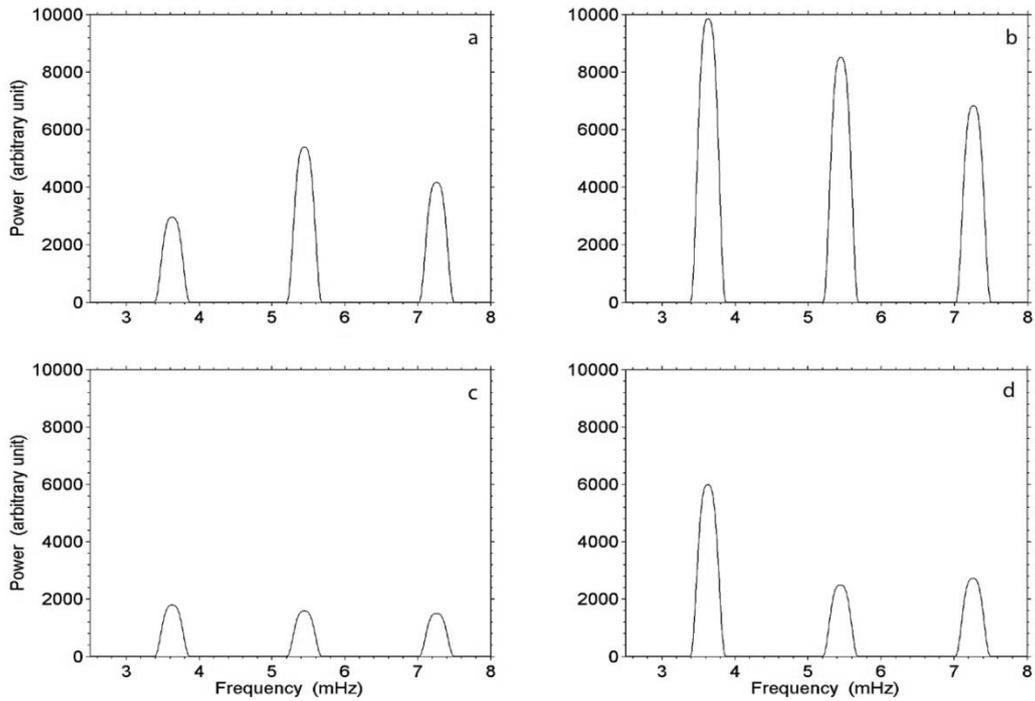

**Fig. 4** Fourier power spectrums obtained on 26 January 2007. The panels a, b, c and d are shown power spectrums, correspond to four separate heights labeled by four red color arrows, respectively, in figure 2 panel a.

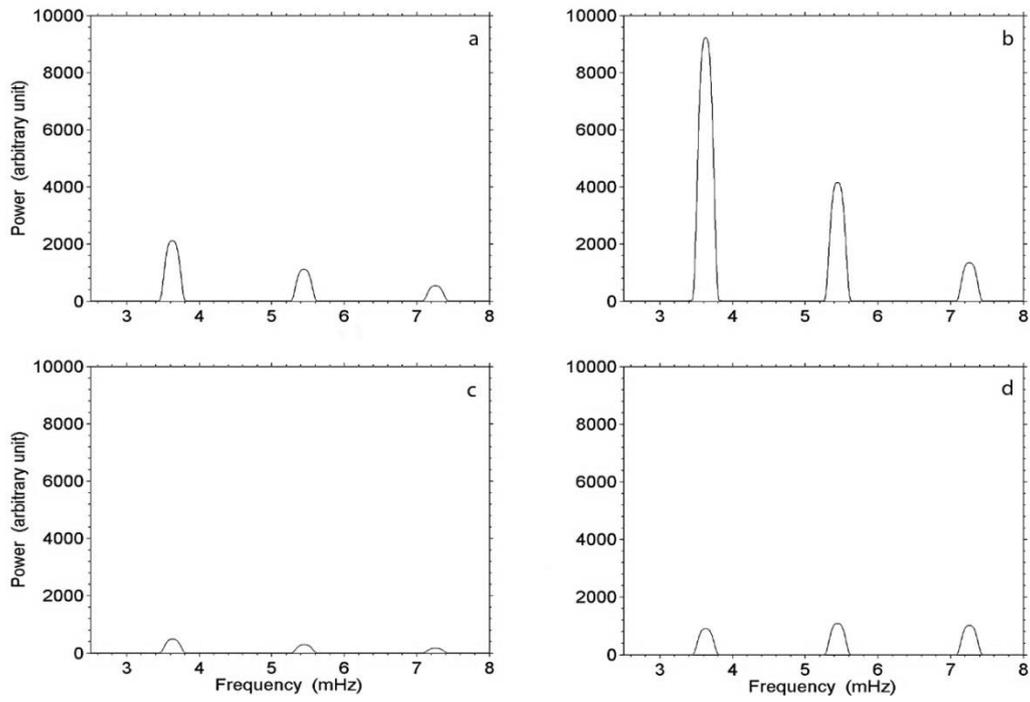

**Fig. 5** Fourier power spectrums obtained on 15 January 2014. The panels a, b, c and d are shown power spectrums, correspond to four separate heights labeled by four red color arrows, respectively, in figure 2 panel a.

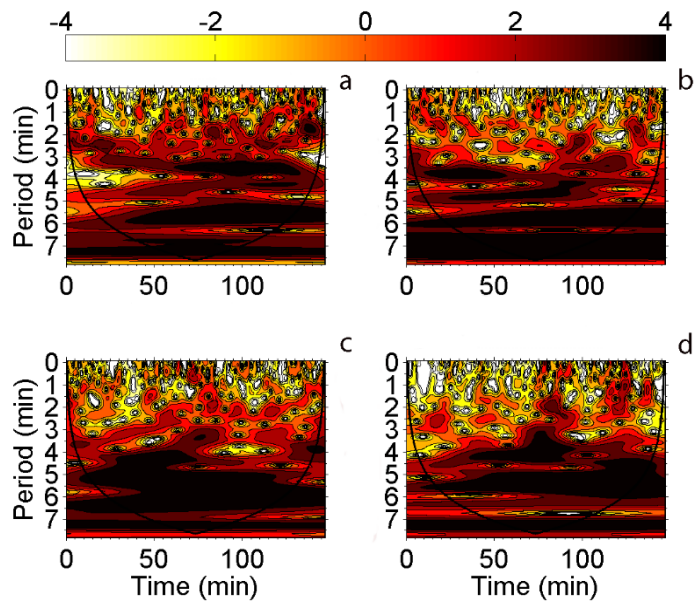

**Fig. 6** The wavelet results for four separate heights obtained from intensity profiles for 22 November 2006. Panels a, b, c and d represent wavelet results corresponding to four separate heights are shown by four red color arrows in figure 2 panel a. The solid black curve line on the panels indicate the cone of influence region (COI) where the wavelet power spectra are distorted because of the influence of the end points of finite length signals.

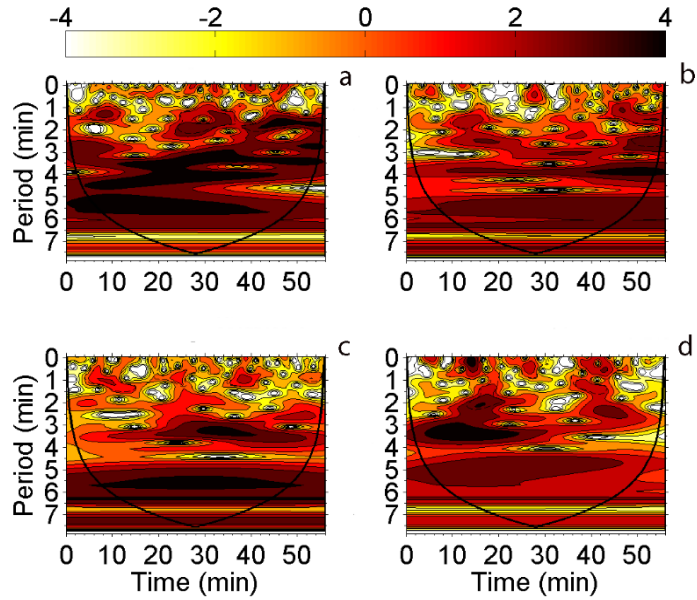

**Fig. 7** The wavelet results for four separate heights obtained from intensity profiles for 26 January 2007. Panels a, b, c and d represent wavelet results corresponding to four separate heights are shown by four red color arrows in figure 2 panel a. The solid black curve line on the panels indicate the cone of influence region (COI) where the wavelet power spectra are distorted because of the influence of the end points of finite length signals.

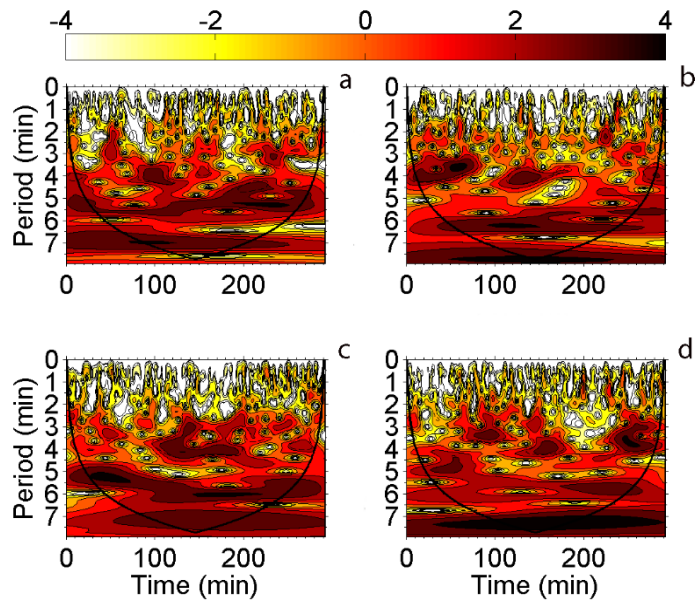

**Fig. 8** The wavelet results for four separate heights obtained from intensity profiles for 15 January 2014. Panels a, b, c and d represent wavelet results corresponding to four separate heights are shown by four red color arrows in figure 2 panel a. The solid black curve line on the panels indicate the cone of influence region (COI) where the wavelet power spectra are distorted because of the influence of the end points of finite length signals.

The global wavelet plots are obtained by taking the mean over the time domain are very similar to the FFT power as both are giving the plots of power with respect to period or frequency. The wavelet results for 22 November 2006 dataset (figure 6) show dominant frequencies about 3 min, 4 min and 5 min similar to FFT results (figure 3) .The wavelet results for two other sets (figures 7 and 8) show dominant frequency about 2 min to 5 min similar to FFT results (figures 4 and 5). Zaqarashvili & Erdélyi (2009) and Gupta et al. (2013) have obtained similar results.

The solid black curve line on the panels of figures 6, 7, 8 indicate the cone of influence region (COI) where the wavelet power spectra are distorted because of the influence of the end points of finite length signals.

### 4.2. Phase difference analysis and phase speed between two certain heights

Comparing the results of FFT and wavelet analysis, we see similar oscillations in all time sequences at four separate heights. It seems that the disturbances producing these oscillations could be due to the waves propagating between two heights. To investigate whether the waves are coherence, we perform FFT and wavelet software package at different heights and calculate the coherency of oscillations first and second, second and third, third and fourth heights, correspond to the left, middle and right panels in figures 9, 10 and 11. (The first, second, third and fourth heights are shown by four red color arrows on figure 2 panel a). The results preserving the distribution of the phase difference represent coherency at frequency about 3.5 mHz and 5.5 mHz.

We measured the coherency and phase difference between two waves at two certain heights as a function of frequency (Trauth 2007 and Oppenheim & Verghese 2010) to obtain phase speed. These calculations are based on cross spectrum phase and coherence estimates for maximum coherency. The coherency results versus frequency are shown for three datasets in figures 12, 13 and 14. The left, middle and right panels show the coherency between two heights, i.e., first and second, second and third, third and fourth heights, respectively. The four mentioned heights are shown in figure 2 panel a. We also calculated phase differences over the full −180° to +180° (360°) range and as a function of the measured oscillation frequency for all heights mentioned above.

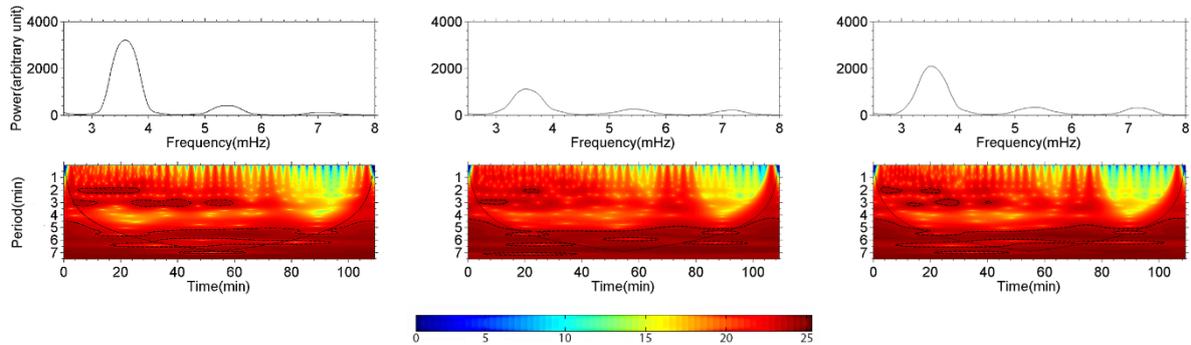

**Fig. 9** Fourier power and wavelet spectrum of 22 November 2006 dataset results for coherency of two certain heights. The left, middle and right Panels show the Fourier coherency power and wavelet coherency spectrum between two certain heights i.e., first and second, second and third, third and fourth heights, respectively. The heights are shown by four red color arrows in figure 2 panel a.

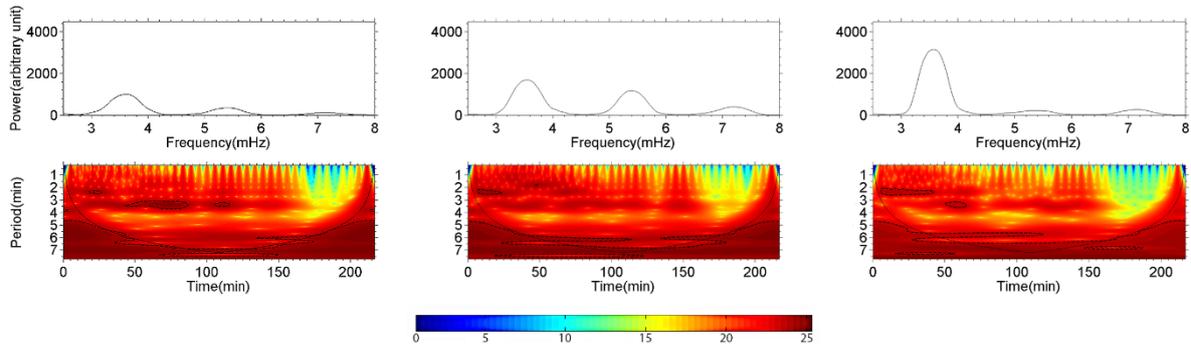

**Fig. 10** Fourier power and wavelet spectrum of 26 January 2007 dataset results for coherency of two certain heights. The left, middle and right panels show the Fourier coherency power and wavelet coherency spectrum between two certain heights i.e., first and second, second and third, third and fourth heights, respectively. The heights are shown by four red color arrows in figure 2 panel a.

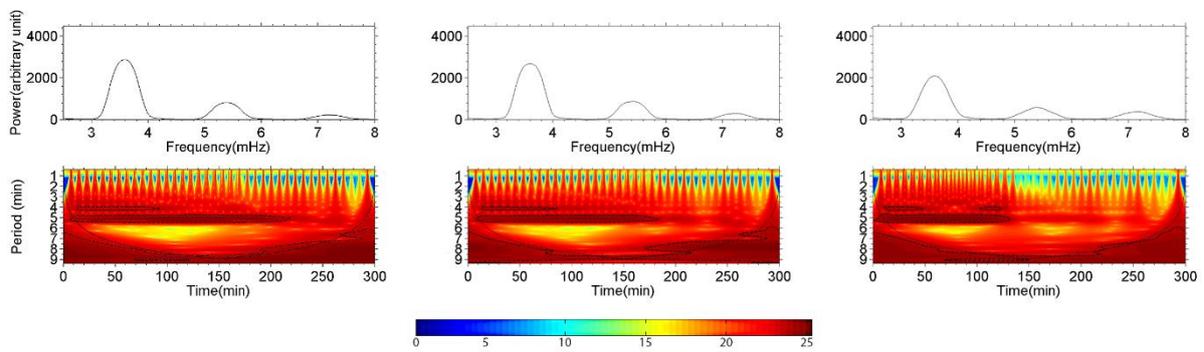

**Fig. 11** Fourier power and wavelet spectrum of 15 January 2014 dataset results for coherency of two certain heights. Left, middle and right panels show the Fourier coherency power and wavelet coherency spectrum between two certain heights i.e., first and second, second and third, third and fourth heights, respectively. The heights are shown by four red color arrows in figure 2 panel a.

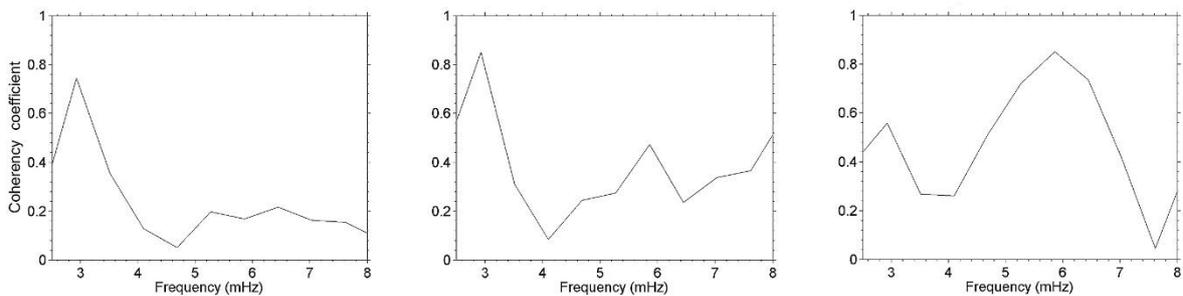

**Fig. 12** Coherency plots for 22 November 2006 dataset. The plots show coherency of waves propagated among different heights. The left, middle and right panels show coherency between two heights i.e., first and second, second and third, third and fourth heights, respectively. The heights are shown by four red color arrows in figure 2 panel a.

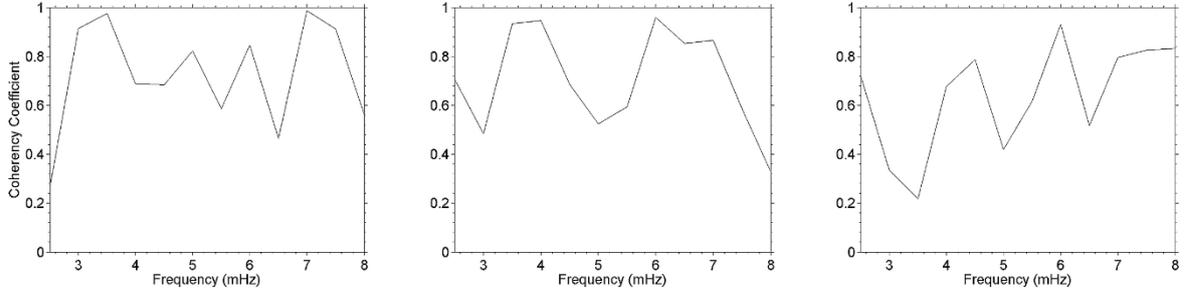

**Fig. 13** Coherency plots for 26 January 2007 dataset. The plots show coherency of waves propagated among different heights. The left, middle and right panels show coherency between two heights i.e., first and second, second and third, third and fourth heights, respectively. The heights are shown by four red color arrows in figure 2 panel a.

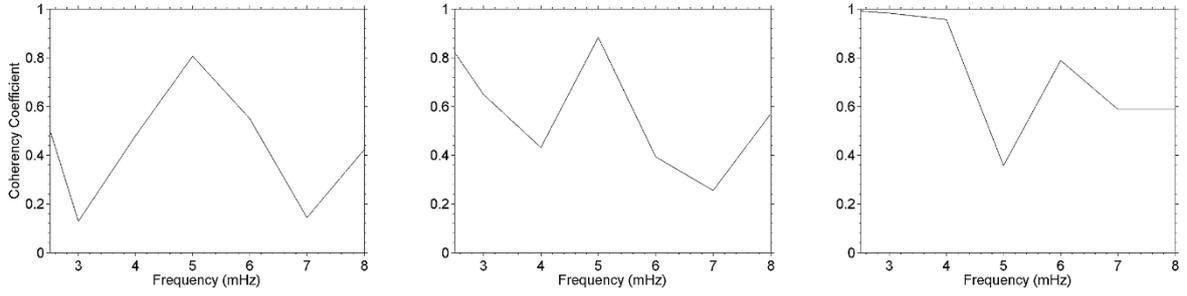

**Fig. 14** Coherency plots for 15 January 2014 dataset. The plots show coherency of waves propagated among different heights. The left, middle and right panels show coherency between two heights i.e., first and second, second and third, third and fourth heights, respectively. The heights are shown by 1, 2, 3, and 4 red color arrows in figure 2 panel a.

By knowing the frequency and phase difference for maximum coherency, we can calculate time difference between two wave propagated phases between two certain heights regarding to equation of phase difference as a function of frequency:

$$\Delta\varphi = 2\pi f\, T$$

Where f is the frequency and T is the time difference in seconds. Then from equation $V_{ph} = H/T$ (the H is the distance between two heights), we can calculate the phase speed.

According to the above expression (measured phase difference and the phase speed of the largest amount of coherency), we can calculate phase difference and phase speed for two maximum coherency obtained from all of coherency plots in each dataset. In this way we can take in to account all the results of time-slices in each dataset. Figure 15, 16, 17 indicate histograms of the frequency for the first and second peaks of coherency for all images in the three dataset.

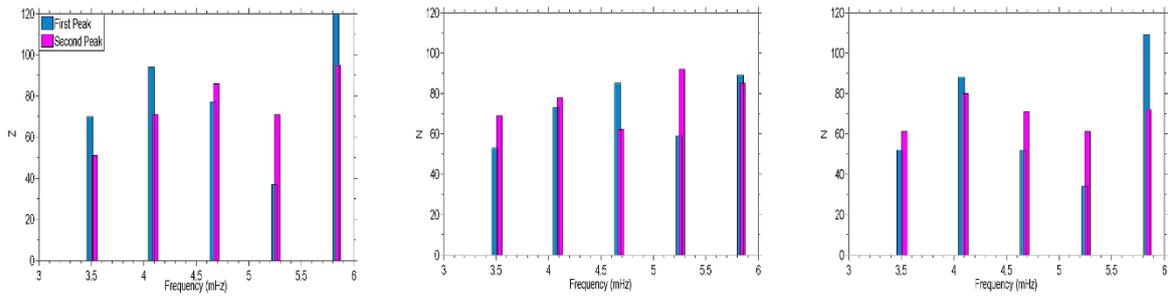

**Fig. 15** Histograms of frequency that shows the coherency of oscillations with similar frequencies for the first and second peaks of coherency for all images in the 22 November 2006 dataset. The left, middle and right are the histograms of frequency for the first and second peaks of coherency between two heights i.e., first and second, second and third, third and fourth heights, respectively. The heights are shown by four red color arrows in figure 2 panel a.

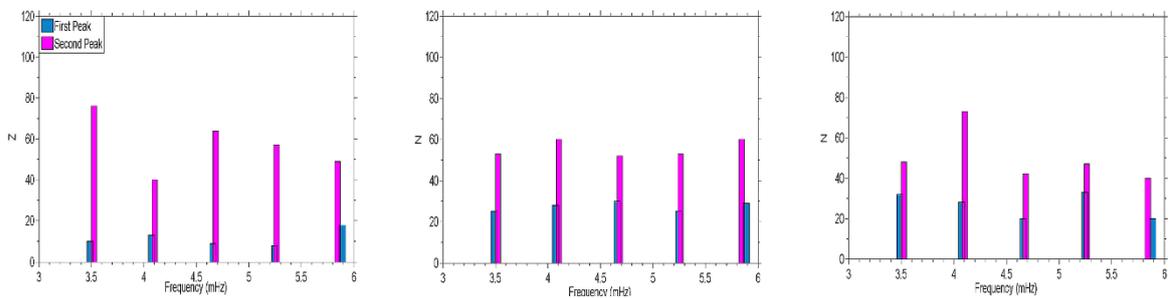

**Fig. 16** Histograms of frequency for the coherency of oscillations with similar frequencies of all images in the 26 January 2007 dataset. The left, middle and right panels are the histograms of frequency for the first and second peaks of coherency between two heights i.e., first and second, second and third, third and fourth heights, respectively. These heights are shown by four red arrows in figure 2 panel a. In the range of defined for histogram, middle and right ones, do not include frequency for the first peaks.

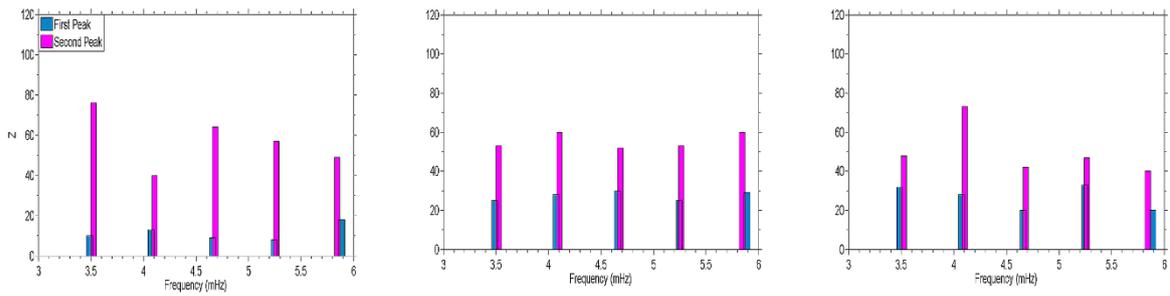

**Fig. 17** Histograms of frequency for the coherency of oscillations with similar frequencies of all images in the 15 January 2014 dataset. The left, middle and right panels are the histograms of frequency for the first and second peaks of coherency between two heights i.e., first and second, second and third, third and fourth heights, respectively. These heights are shown by four red arrows in figure 2 panel a. In the range of defined for histogram, middle and right ones, do not include frequency for the first peaks.

Histograms of frequency for 22 November 2006 dataset (quiet Sun), 26 January 2007 dataset (active region) and 15 January 2014 dataset (active Sun) display frequencies about 3.5 mHz, 4.2 mHz, 4.6 mHz, 5.3 mHz and 5.8 mHz, for the first and second peaks of coherency, see figures 15, 16 and 17.

The results of the phase speeds for the dominant frequencies are given in table 2 for the three datasets.

**Table 2.** Phase speeds for three datasets. Columns a, b and c show phase speed between two heights i.e., first and second, second and third, third and fourth heights, respectively. The heights are shown by four red color arrows in figure 2 panel a.

| Date | Phase speed ± 15 ($kms^{-1}$) | | |
|---|---|---|---|
| | a | b | c |
| 22 November 2006 (quiet Sun) | 50-300 | 100-400 | 150-450 |
| 26 January 2007 (active region) | 50-450 | 150-550 | 200-650 |
| 15 January 2014 (active Sun) | 50-400 | 100-500 | 200-550 |

Result of table 2 shows that, the phase speeds of 50-450 $kms^{-1}$ is measured for quiet Sun, 50-650 $kms^{-1}$ for active region and 50-550 $kms^{-1}$ for active Sun. The majority of the measured phase speeds in locations where there in known to be considerable dynamic activity are more than quiet Sun. It is also seen, the mean phase speeds obtained from three datasets increase with height, see table 2, because the density of environment decreases with height.

## 5. Summery and conclusion

Coherency of solar oscillations were studied using the Fourier power, wavelet spectrum and phase difference analysis in active Sun, quiet Sun and active region from the intensity time series observed by HINODE/SOT in Ca II H line.

The results of FFT plots for the three datasets show frequencies about 3.6 mHz, 5.5 mHz and 7.3 mHz at four separate heights. It can be seen that increase and reduction of powers are not regular at four separate heights for the three datasets, more like randomly behavior. It is noted that the results randomly are obtained for selective time-slices, and the results are consistent which obtained by Hanstoen et al. (2000) and Tavabi (2014). Gupta et al. 2013 studied quiet Sun oscillations using the Fourier power and phase difference analysis from the intensity time series. They found that the Fourier power maps reveal 3 and 5 minutes around the bright magnetic–network regions at chromospheric heights. Jefferies et al. (2006) explained that, the low–frequency photospheric oscillations could propagate into the solar chromosphere through "magneto-acoustic portals". Banerjee et al. (2001) observed 2–4 mHz network oscillations in the low chromospheric and transition region lines in both intensity and velocity, which were interpreted in terms of kink and sausage waves propagating upwards along thin magnetic flux tubes.

The wavelet results for quiet Sun, active Sun and active region indicate dominant frequencies about 3 min, 4 min and 5 min similar to FFT results (Zaqarashvili & Erdélyi 2009 and Gupta et al. 2013). As expected the results coherency at two certain heights represent frequencies at about 3.5 mHz and 5.5 mHz oscillations for three datasets by FFT and wavelet software package.

Histograms of frequency for 22 November 2006 dataset (quiet Sun), 26 January 2007 dataset (active region) and 15 January 2014 dataset (active Sun) display frequencies about 3.5 mHz, 4.2 mHz, 4.6 mHz, 5.3 mHz and 5.8 mHz for the first and second peak of coherency.

The phase speeds of 50-450 $kms^{-1}$ (See table 2) are measured for quiet Sun, 50-650 $kms^{-1}$ for active region and 50-550 $kms^{-1}$ for active Sun (McIntosh et al. 2011). The majority of the measured phase speeds in locations where there is known to be considerable dynamic activity are more than those of quiet Sun. The mean phase speeds obtained from three datasets, increase with height, because the density of environment decreases with height.

In this study, we obtained that these waves are collective and show group behavior, since they are coherent perturbations of all confined magnetical flux inside the spicule. Phase and group speed are determined by the coherency frequencies of the spicule and the flow velocity. These results could be used for modeling of spicules in solar chromosphere.

**Acknowledgements**. We are indebted to Serge Koutchmy for idea and making meaningful remarks on the paper. We are grateful to the HINODE team for their wonderful observations. HINODE is a Japanese mission developed and launched by ISAS/JAXA, with NAOJ as domestic partner and NASA, ESA, and STFC (UK) as international partners. Image processing software was provided by O. Koutchmy, see http:// www .ann .jussieu.fr/ koutchmy/index newE.html. The standard SOT subroutines can be found in the SSWIDL software tree (http://sohowww.nascom.nasa.gov/solarsoft/hinode/sot/idl). The wavelet software package and compilation instructions were downloaded from http://paos.colorado.edu/research/wavelets/software.html. This work has been supported by Research Institute for Astronomy & Astrophysics of Maragha (RIAAM) and the Center for Excellence in Astronomy & Astrophysics (CEAA), and Islamic Azad University, Tabriz Branch (IAUT), and UPMC.